\hsize=31pc
\vsize=49pc
\lineskip=0pt
\parskip=0pt plus 1pt
\hfuzz=1pt  
\vfuzz=2pt
\pretolerance=2500
\tolerance=5000
\vbadness=5000
\hbadness=5000
\widowpenalty=500
\clubpenalty=200
\brokenpenalty=500
\predisplaypenalty=200
\voffset=-1pc
\nopagenumbers     
\catcode`@=11
\newif\ifams
\amsfalse
%
%
%
%
\newfam\bdifam
\newfam\bsyfam
\newfam\bssfam
\newfam\msafam
\newfam\msbfam
\newif\ifxxpt   
\newif\ifxviipt 
\newif\ifxivpt  
\newif\ifxiipt  
\newif\ifxipt   
\newif\ifxpt    
\newif\ifixpt   
\newif\ifviiipt 
\newif\ifviipt  
\newif\ifvipt   
\newif\ifvpt    
%
%
\def\headsize#1#2{\def\headb@seline{#2}%
                \ifnum#1=20\def\HEAD{twenty}%
                           \def\smHEAD{twelve}%
                           \def\vsHEAD{nine}%
                           \ifxxpt\else\xdef\f@ntsize{\HEAD}%
                           \def\m@g{4}\def\s@ze{20.74}%
                           \loadheadfonts\xxpttrue\fi
                           \ifxiipt\else\xdef\f@ntsize{\smHEAD}%
                           \def\m@g{1}\def\s@ze{12}%
                           \loadxiiptfonts\xiipttrue\fi
                           \ifixpt\else\xdef\f@ntsize{\vsHEAD}%
                           \def\s@ze{9}%
                           \loadsmallfonts\ixpttrue\fi
                      \else
                \ifnum#1=17\def\HEAD{seventeen}%
                           \def\smHEAD{eleven}%
                           \def\vsHEAD{eight}%
                           \ifxviipt\else\xdef\f@ntsize{\HEAD}%
                           \def\m@g{3}\def\s@ze{17.28}%
                           \loadheadfonts\xviipttrue\fi
                           \ifxipt\else\xdef\f@ntsize{\smHEAD}%
                           \loadxiptfonts\xipttrue\fi
                           \ifviiipt\else\xdef\f@ntsize{\vsHEAD}%
                           \def\s@ze{8}%
                           \loadsmallfonts\viiipttrue\fi
                      \else\def\HEAD{fourteen}%
                           \def\smHEAD{ten}%
                           \def\vsHEAD{seven}%
                           \ifxivpt\else\xdef\f@ntsize{\HEAD}%
                           \def\m@g{2}\def\s@ze{14.4}%
                           \loadheadfonts\xivpttrue\fi
                           \ifxpt\else\xdef\f@ntsize{\smHEAD}%
                           \def\s@ze{10}%
                           \loadxptfonts\xpttrue\fi
                           \ifviipt\else\xdef\f@ntsize{\vsHEAD}%
                           \def\s@ze{7}%
                           \loadviiptfonts\viipttrue\fi
                \ifnum#1=14\else
                \message{Header size should be 20, 17 or 14 point
                              will now default to 14pt}\fi
                \fi\fi\headfonts}
%
%
\def\textsize#1#2{\def\textb@seline{#2}%
                 \ifnum#1=12\def\TEXT{twelve}%
                           \def\smTEXT{eight}%
                           \def\vsTEXT{six}%
                           \ifxiipt\else\xdef\f@ntsize{\TEXT}%
                           \def\m@g{1}\def\s@ze{12}%
                           \loadxiiptfonts\xiipttrue\fi
                           \ifviiipt\else\xdef\f@ntsize{\smTEXT}%
                           \def\s@ze{8}%
                           \loadsmallfonts\viiipttrue\fi
                           \ifvipt\else\xdef\f@ntsize{\vsTEXT}%
                           \def\s@ze{6}%
                           \loadviptfonts\vipttrue\fi
                      \else
                \ifnum#1=11\def\TEXT{eleven}%
                           \def\smTEXT{seven}%
                           \def\vsTEXT{five}%
                           \ifxipt\else\xdef\f@ntsize{\TEXT}%
                           \def\s@ze{11}%
                           \loadxiptfonts\xipttrue\fi
                           \ifviipt\else\xdef\f@ntsize{\smTEXT}%
                           \loadviiptfonts\viipttrue\fi
                           \ifvpt\else\xdef\f@ntsize{\vsTEXT}%
                           \def\s@ze{5}%
                           \loadvptfonts\vpttrue\fi
                      \else\def\TEXT{ten}%
                           \def\smTEXT{seven}%
                           \def\vsTEXT{five}%
                           \ifxpt\else\xdef\f@ntsize{\TEXT}%
                           \loadxptfonts\xpttrue\fi
                           \ifviipt\else\xdef\f@ntsize{\smTEXT}%
                           \def\s@ze{7}%
                           \loadviiptfonts\viipttrue\fi
                           \ifvpt\else\xdef\f@ntsize{\vsTEXT}%
                           \def\s@ze{5}%
                           \loadvptfonts\vpttrue\fi
                \ifnum#1=10\else
                \message{Text size should be 12, 11 or 10 point
                              will now default to 10pt}\fi
                \fi\fi\textfonts}
%
%
\def\smallsize#1#2{\def\smallb@seline{#2}%
                 \ifnum#1=10\def\SMALL{ten}%
                           \def\smSMALL{seven}%
                           \def\vsSMALL{five}%
                           \ifxpt\else\xdef\f@ntsize{\SMALL}%
                           \loadxptfonts\xpttrue\fi
                           \ifviipt\else\xdef\f@ntsize{\smSMALL}%
                           \def\s@ze{7}%
                           \loadviiptfonts\viipttrue\fi
                           \ifvpt\else\xdef\f@ntsize{\vsSMALL}%
                           \def\s@ze{5}%
                           \loadvptfonts\vpttrue\fi
                       \else
                 \ifnum#1=9\def\SMALL{nine}%
                           \def\smSMALL{six}%
                           \def\vsSMALL{five}%
                           \ifixpt\else\xdef\f@ntsize{\SMALL}%
                           \def\s@ze{9}%
                           \loadsmallfonts\ixpttrue\fi
                           \ifvipt\else\xdef\f@ntsize{\smSMALL}%
                           \def\s@ze{6}%
                           \loadviptfonts\vipttrue\fi
                           \ifvpt\else\xdef\f@ntsize{\vsSMALL}%
                           \def\s@ze{5}%
                           \loadvptfonts\vpttrue\fi
                       \else
                           \def\SMALL{eight}%
                           \def\smSMALL{six}%
                           \def\vsSMALL{five}%
                           \ifviiipt\else\xdef\f@ntsize{\SMALL}%
                           \def\s@ze{8}%
                           \loadsmallfonts\viiipttrue\fi
                           \ifvipt\else\xdef\f@ntsize{\smSMALL}%
                           \def\s@ze{6}%
                           \loadviptfonts\vipttrue\fi
                           \ifvpt\else\xdef\f@ntsize{\vsSMALL}%
                           \def\s@ze{5}%
                           \loadvptfonts\vpttrue\fi
                 \ifnum#1=8\else\message{Small size should be 10, 9 or 
                            8 point will now default to 8pt}\fi
                \fi\fi\smallfonts}
\def\F@nt{\expandafter\font\csname}
\def\Sk@w{\expandafter\skewchar\csname}
\def\@nd{\endcsname}
\def\@step#1{ scaled \magstep#1}
\def\@half{ scaled \magstephalf}
\def\@t#1{ at #1pt}
%
%
\def\loadheadfonts{\bigf@nts
\F@nt \f@ntsize bdi\@nd=cmmib10 \@t{\s@ze}%
\Sk@w \f@ntsize bdi\@nd='177
\F@nt \f@ntsize bsy\@nd=cmbsy10 \@t{\s@ze}%
\Sk@w \f@ntsize bsy\@nd='60
\F@nt \f@ntsize bss\@nd=cmssbx10 \@t{\s@ze}}
%
%
\def\loadxiiptfonts{\bigf@nts
\F@nt \f@ntsize bdi\@nd=cmmib10 \@step{\m@g}%
\Sk@w \f@ntsize bdi\@nd='177
\F@nt \f@ntsize bsy\@nd=cmbsy10 \@step{\m@g}%
\Sk@w \f@ntsize bsy\@nd='60
\F@nt \f@ntsize bss\@nd=cmssbx10 \@step{\m@g}}
%
%
\def\loadxiptfonts{%
\font\elevenrm=cmr10 \@half
\font\eleveni=cmmi10 \@half
\skewchar\eleveni='177
\font\elevensy=cmsy10 \@half
\skewchar\elevensy='60
\font\elevenex=cmex10 \@half
\font\elevenit=cmti10 \@half
\font\elevensl=cmsl10 \@half
\font\elevenbf=cmbx10 \@half
\font\eleventt=cmtt10 \@half
\ifams\font\elevenmsa=msam10 \@half
\font\elevenmsb=msbm10 \@half\else\fi
\font\elevenbdi=cmmib10 \@half
\skewchar\elevenbdi='177
\font\elevenbsy=cmbsy10 \@half
\skewchar\elevenbsy='60
\font\elevenbss=cmssbx10 \@half}
%
%
\def\loadxptfonts{%
\font\tenbdi=cmmib10
\skewchar\tenbdi='177
\font\tenbsy=cmbsy10 
\skewchar\tenbsy='60
\ifams\font\tenmsa=msam10 
\font\tenmsb=msbm10\else\fi
\font\tenbss=cmssbx10}%
%
%
\def\loadsmallfonts{\smallf@nts
\ifams
\F@nt \f@ntsize ex\@nd=cmex\s@ze
\else
\F@nt \f@ntsize ex\@nd=cmex10\fi
\F@nt \f@ntsize it\@nd=cmti\s@ze
\F@nt \f@ntsize sl\@nd=cmsl\s@ze
\F@nt \f@ntsize tt\@nd=cmtt\s@ze}
%
%
\def\loadviiptfonts{%
\font\sevenit=cmti7
\font\sevensl=cmsl8 at 7pt
\ifams\font\sevenmsa=msam7 
\font\sevenmsb=msbm7
\font\sevenex=cmex7
\font\sevenbsy=cmbsy7
\font\sevenbdi=cmmib7\else
\font\sevenex=cmex10
\font\sevenbsy=cmbsy10 at 7pt
\font\sevenbdi=cmmib10 at 7pt\fi
\skewchar\sevenbsy='60
\skewchar\sevenbdi='177
\font\sevenbss=cmssbx10 at 7pt}%
%
%
\def\loadviptfonts{\smallf@nts
\ifams\font\sixex=cmex7 at 6pt\else
\font\sixex=cmex10\fi
\font\sixit=cmti7 at 6pt}
%
%
\def\loadvptfonts{%
\font\fiveit=cmti7 at 5pt
\ifams\font\fiveex=cmex7 at 5pt
\font\fivebdi=cmmib5
\font\fivebsy=cmbsy5
\font\fivemsa=msam5 
\font\fivemsb=msbm5\else
\font\fiveex=cmex10
\font\fivebdi=cmmib10 at 5pt
\font\fivebsy=cmbsy10 at 5pt\fi
\skewchar\fivebdi='177
\skewchar\fivebsy='60
\font\fivebss=cmssbx10 at 5pt}
\def\bigf@nts{%
\F@nt \f@ntsize rm\@nd=cmr10 \@step{\m@g}%
\F@nt \f@ntsize i\@nd=cmmi10 \@step{\m@g}%
\Sk@w \f@ntsize i\@nd='177
\F@nt \f@ntsize sy\@nd=cmsy10 \@step{\m@g}%
\Sk@w \f@ntsize sy\@nd='60
\F@nt \f@ntsize ex\@nd=cmex10 \@step{\m@g}%
\F@nt \f@ntsize it\@nd=cmti10 \@step{\m@g}%
\F@nt \f@ntsize sl\@nd=cmsl10 \@step{\m@g}%
\F@nt \f@ntsize bf\@nd=cmbx10 \@step{\m@g}%
\F@nt \f@ntsize tt\@nd=cmtt10 \@step{\m@g}%
\ifams
\F@nt \f@ntsize msa\@nd=msam10 \@step{\m@g}%
\F@nt \f@ntsize msb\@nd=msbm10 \@step{\m@g}\else\fi}
\def\smallf@nts{%
\F@nt \f@ntsize rm\@nd=cmr\s@ze
\F@nt \f@ntsize i\@nd=cmmi\s@ze 
\Sk@w \f@ntsize i\@nd='177
\F@nt \f@ntsize sy\@nd=cmsy\s@ze
\Sk@w \f@ntsize sy\@nd='60
\F@nt \f@ntsize bf\@nd=cmbx\s@ze 
\ifams
\F@nt \f@ntsize bdi\@nd=cmmib\s@ze 
\F@nt \f@ntsize bsy\@nd=cmbsy\s@ze 
\F@nt \f@ntsize msa\@nd=msam\s@ze 
\F@nt \f@ntsize msb\@nd=msbm\s@ze
\else
\F@nt \f@ntsize bdi\@nd=cmmib10 \@t{\s@ze}%
\F@nt \f@ntsize bsy\@nd=cmbsy10 \@t{\s@ze}\fi 
\Sk@w \f@ntsize bdi\@nd='177
\Sk@w \f@ntsize bsy\@nd='60
\F@nt \f@ntsize bss\@nd=cmssbx10 \@t{\s@ze}}%
%
%
\def\headfonts{%
\textfont0=\csname\HEAD rm\@nd        
\scriptfont0=\csname\smHEAD rm\@nd
\scriptscriptfont0=\csname\vsHEAD rm\@nd
\def\rm{\fam0\csname\HEAD rm\@nd
\def\sc{\csname\smHEAD rm\@nd}}%
\textfont1=\csname\HEAD i\@nd         
\scriptfont1=\csname\smHEAD i\@nd
\scriptscriptfont1=\csname\vsHEAD i\@nd
\textfont2=\csname\HEAD sy\@nd        
\scriptfont2=\csname\smHEAD sy\@nd
\scriptscriptfont2=\csname\vsHEAD sy\@nd
\textfont3=\csname\HEAD ex\@nd        
\scriptfont3=\csname\smHEAD ex\@nd
\scriptscriptfont3=\csname\smHEAD ex\@nd
\textfont\itfam=\csname\HEAD it\@nd   
\scriptfont\itfam=\csname\smHEAD it\@nd
\scriptscriptfont\itfam=\csname\vsHEAD it\@nd
\def\it{\fam\itfam\csname\HEAD it\@nd
\def\sc{\csname\smHEAD it\@nd}}%
\textfont\slfam=\csname\HEAD sl\@nd   
\def\sl{\fam\slfam\csname\HEAD sl\@nd
\def\sc{\csname\smHEAD sl\@nd}}%
\textfont\bffam=\csname\HEAD bf\@nd   
\scriptfont\bffam=\csname\smHEAD bf\@nd
\scriptscriptfont\bffam=\csname\vsHEAD bf\@nd
\def\bf{\fam\bffam\csname\HEAD bf\@nd
\def\sc{\csname\smHEAD bf\@nd}}%
\textfont\ttfam=\csname\HEAD tt\@nd   
\def\tt{\fam\ttfam\csname\HEAD tt\@nd}%
\textfont\bdifam=\csname\HEAD bdi\@nd 
\scriptfont\bdifam=\csname\smHEAD bdi\@nd
\scriptscriptfont\bdifam=\csname\vsHEAD bdi\@nd
\def\bdi{\fam\bdifam\csname\HEAD bdi\@nd}%
\textfont\bsyfam=\csname\HEAD bsy\@nd 
\scriptfont\bsyfam=\csname\smHEAD bsy\@nd
\def\bsy{\fam\bsyfam\csname\HEAD bsy\@nd}%
\textfont\bssfam=\csname\HEAD bss\@nd 
\scriptfont\bssfam=\csname\smHEAD bss\@nd
\scriptscriptfont\bssfam=\csname\vsHEAD bss\@nd
\def\bss{\fam\bssfam\csname\HEAD bss\@nd}%
\ifams
\textfont\msafam=\csname\HEAD msa\@nd 
\scriptfont\msafam=\csname\smHEAD msa\@nd
\scriptscriptfont\msafam=\csname\vsHEAD msa\@nd
\textfont\msbfam=\csname\HEAD msb\@nd 
\scriptfont\msbfam=\csname\smHEAD msb\@nd
\scriptscriptfont\msbfam=\csname\vsHEAD msb\@nd
\else\fi
\normalbaselineskip=\headb@seline pt%
\setbox\strutbox=\hbox{\vrule height.7\normalbaselineskip 
depth.3\baselineskip width0pt}%
\def\sc{\csname\smHEAD rm\@nd}\normalbaselines\bf}
%
%
\def\textfonts{%
\textfont0=\csname\TEXT rm\@nd        
\scriptfont0=\csname\smTEXT rm\@nd
\scriptscriptfont0=\csname\vsTEXT rm\@nd
\def\rm{\fam0\csname\TEXT rm\@nd
\def\sc{\csname\smTEXT rm\@nd}}%
\textfont1=\csname\TEXT i\@nd         
\scriptfont1=\csname\smTEXT i\@nd
\scriptscriptfont1=\csname\vsTEXT i\@nd
\textfont2=\csname\TEXT sy\@nd        
\scriptfont2=\csname\smTEXT sy\@nd
\scriptscriptfont2=\csname\vsTEXT sy\@nd
\textfont3=\csname\TEXT ex\@nd        
\scriptfont3=\csname\smTEXT ex\@nd
\scriptscriptfont3=\csname\smTEXT ex\@nd
\textfont\itfam=\csname\TEXT it\@nd   
\scriptfont\itfam=\csname\smTEXT it\@nd
\scriptscriptfont\itfam=\csname\vsTEXT it\@nd
\def\it{\fam\itfam\csname\TEXT it\@nd
\def\sc{\csname\smTEXT it\@nd}}%
\textfont\slfam=\csname\TEXT sl\@nd   
\def\sl{\fam\slfam\csname\TEXT sl\@nd
\def\sc{\csname\smTEXT sl\@nd}}%
\textfont\bffam=\csname\TEXT bf\@nd   
\scriptfont\bffam=\csname\smTEXT bf\@nd
\scriptscriptfont\bffam=\csname\vsTEXT bf\@nd
\def\bf{\fam\bffam\csname\TEXT bf\@nd
\def\sc{\csname\smTEXT bf\@nd}}%
\textfont\ttfam=\csname\TEXT tt\@nd   
\def\tt{\fam\ttfam\csname\TEXT tt\@nd}%
\textfont\bdifam=\csname\TEXT bdi\@nd 
\scriptfont\bdifam=\csname\smTEXT bdi\@nd
\scriptscriptfont\bdifam=\csname\vsTEXT bdi\@nd
\def\bdi{\fam\bdifam\csname\TEXT bdi\@nd}%
\textfont\bsyfam=\csname\TEXT bsy\@nd 
\scriptfont\bsyfam=\csname\smTEXT bsy\@nd
\def\bsy{\fam\bsyfam\csname\TEXT bsy\@nd}%
\textfont\bssfam=\csname\TEXT bss\@nd 
\scriptfont\bssfam=\csname\smTEXT bss\@nd
\scriptscriptfont\bssfam=\csname\vsTEXT bss\@nd
\def\bss{\fam\bssfam\csname\TEXT bss\@nd}%
\ifams
\textfont\msafam=\csname\TEXT msa\@nd 
\scriptfont\msafam=\csname\smTEXT msa\@nd
\scriptscriptfont\msafam=\csname\vsTEXT msa\@nd
\textfont\msbfam=\csname\TEXT msb\@nd 
\scriptfont\msbfam=\csname\smTEXT msb\@nd
\scriptscriptfont\msbfam=\csname\vsTEXT msb\@nd
\else\fi
\normalbaselineskip=\textb@seline pt
\setbox\strutbox=\hbox{\vrule height.7\normalbaselineskip 
depth.3\baselineskip width0pt}%
\everymath{}%
\def\sc{\csname\smTEXT rm\@nd}\normalbaselines\rm}
%
%
\def\smallfonts{%
\textfont0=\csname\SMALL rm\@nd        
\scriptfont0=\csname\smSMALL rm\@nd
\scriptscriptfont0=\csname\vsSMALL rm\@nd
\def\rm{\fam0\csname\SMALL rm\@nd
\def\sc{\csname\smSMALL rm\@nd}}%
\textfont1=\csname\SMALL i\@nd         
\scriptfont1=\csname\smSMALL i\@nd
\scriptscriptfont1=\csname\vsSMALL i\@nd
\textfont2=\csname\SMALL sy\@nd        
\scriptfont2=\csname\smSMALL sy\@nd
\scriptscriptfont2=\csname\vsSMALL sy\@nd
\textfont3=\csname\SMALL ex\@nd        
\scriptfont3=\csname\smSMALL ex\@nd
\scriptscriptfont3=\csname\smSMALL ex\@nd
\textfont\itfam=\csname\SMALL it\@nd   
\scriptfont\itfam=\csname\smSMALL it\@nd
\scriptscriptfont\itfam=\csname\vsSMALL it\@nd
\def\it{\fam\itfam\csname\SMALL it\@nd
\def\sc{\csname\smSMALL it\@nd}}%
\textfont\slfam=\csname\SMALL sl\@nd   
\def\sl{\fam\slfam\csname\SMALL sl\@nd
\def\sc{\csname\smSMALL sl\@nd}}%
\textfont\bffam=\csname\SMALL bf\@nd   
\scriptfont\bffam=\csname\smSMALL bf\@nd
\scriptscriptfont\bffam=\csname\vsSMALL bf\@nd
\def\bf{\fam\bffam\csname\SMALL bf\@nd
\def\sc{\csname\smSMALL bf\@nd}}%
\textfont\ttfam=\csname\SMALL tt\@nd   
\def\tt{\fam\ttfam\csname\SMALL tt\@nd}%
\textfont\bdifam=\csname\SMALL bdi\@nd 
\scriptfont\bdifam=\csname\smSMALL bdi\@nd
\scriptscriptfont\bdifam=\csname\vsSMALL bdi\@nd
\def\bdi{\fam\bdifam\csname\SMALL bdi\@nd}%
\textfont\bsyfam=\csname\SMALL bsy\@nd 
\scriptfont\bsyfam=\csname\smSMALL bsy\@nd
\def\bsy{\fam\bsyfam\csname\SMALL bsy\@nd}%
\textfont\bssfam=\csname\SMALL bss\@nd 
\scriptfont\bssfam=\csname\smSMALL bss\@nd
\scriptscriptfont\bssfam=\csname\vsSMALL bss\@nd
\def\bss{\fam\bssfam\csname\SMALL bss\@nd}%
\ifams
\textfont\msafam=\csname\SMALL msa\@nd 
\scriptfont\msafam=\csname\smSMALL msa\@nd
\scriptscriptfont\msafam=\csname\vsSMALL msa\@nd
\textfont\msbfam=\csname\SMALL msb\@nd 
\scriptfont\msbfam=\csname\smSMALL msb\@nd
\scriptscriptfont\msbfam=\csname\vsSMALL msb\@nd
\else\fi
\normalbaselineskip=\smallb@seline pt%
\setbox\strutbox=\hbox{\vrule height.7\normalbaselineskip 
depth.3\baselineskip width0pt}%
\everymath{}%
\def\sc{\csname\smSMALL rm\@nd}\normalbaselines\rm}%
\everydisplay{\indenteddisplay
   \gdef\labeltype{\eqlabel}}%
%
%
\def\hexnumber@#1{\ifcase#1 0\or 1\or 2\or 3\or 4\or 5\or 6\or 7\or 8\or
 9\or A\or B\or C\or D\or E\or F\fi}
\edef\bffam@{\hexnumber@\bffam}
\edef\bdifam@{\hexnumber@\bdifam}
\edef\bsyfam@{\hexnumber@\bsyfam}
\def\undefine#1{\let#1\undefined}
\def\newsymbol#1#2#3#4#5{\let\next@\relax
 \ifnum#2=\thr@@\let\next@\bdifam@\else
 \ifams
 \ifnum#2=\@ne\let\next@\msafam@\else
 \ifnum#2=\tw@\let\next@\msbfam@\fi\fi
 \fi\fi
 \mathchardef#1="#3\next@#4#5}
\def\mathhexbox@#1#2#3{\relax
 \ifmmode\mathpalette{}{\m@th\mathchar"#1#2#3}%
 \else\leavevmode\hbox{$\m@th\mathchar"#1#2#3$}\fi}

\def\bi#1{{\fam\bdifam\relax#1}}
%
%
\ifams\input amsmacro\fi
%
%
\newsymbol\bitGamma 3000
\newsymbol\bitDelta 3001
\newsymbol\bitTheta 3002
\newsymbol\bitLambda 3003
\newsymbol\bitXi 3004
\newsymbol\bitPi 3005
\newsymbol\bitSigma 3006
\newsymbol\bitUpsilon 3007
\newsymbol\bitPhi 3008
\newsymbol\bitPsi 3009
\newsymbol\bitOmega 300A
\newsymbol\balpha 300B
\newsymbol\bbeta 300C
\newsymbol\bgamma 300D
\newsymbol\bdelta 300E
\newsymbol\bepsilon 300F
\newsymbol\bzeta 3010
\newsymbol\bfeta 3011
\newsymbol\btheta 3012
\newsymbol\biota 3013
\newsymbol\bkappa 3014
\newsymbol\blambda 3015
\newsymbol\bmu 3016
\newsymbol\bnu 3017
\newsymbol\bxi 3018
\newsymbol\bpi 3019
\newsymbol\brho 301A
\newsymbol\bsigma 301B
\newsymbol\btau 301C
\newsymbol\bupsilon 301D
\newsymbol\bphi 301E
\newsymbol\bchi 301F
\newsymbol\bpsi 3020
\newsymbol\bomega 3021
\newsymbol\bvarepsilon 3022
\newsymbol\bvartheta 3023
\newsymbol\bvaromega 3024
\newsymbol\bvarrho 3025
\newsymbol\bvarzeta 3026
\newsymbol\bvarphi 3027
\newsymbol\bpartial 3040
\newsymbol\bell 3060
\newsymbol\bimath 307B
\newsymbol\bjmath 307C
\mathchardef\binfty "0\bsyfam@31
\mathchardef\bnabla "0\bsyfam@72
\mathchardef\bdot "2\bsyfam@01
\mathchardef\bGamma "0\bffam@00
\mathchardef\bDelta "0\bffam@01
\mathchardef\bTheta "0\bffam@02
\mathchardef\bLambda "0\bffam@03
\mathchardef\bXi "0\bffam@04
\mathchardef\bPi "0\bffam@05
\mathchardef\bSigma "0\bffam@06
\mathchardef\bUpsilon "0\bffam@07
\mathchardef\bPhi "0\bffam@08
\mathchardef\bPsi "0\bffam@09
\mathchardef\bOmega "0\bffam@0A
\mathchardef\itGamma "0100
\mathchardef\itDelta "0101
\mathchardef\itTheta "0102
\mathchardef\itLambda "0103
\mathchardef\itXi "0104
\mathchardef\itPi "0105
\mathchardef\itSigma "0106
\mathchardef\itUpsilon "0107
\mathchardef\itPhi "0108
\mathchardef\itPsi "0109
\mathchardef\itOmega "010A
\mathchardef\Gamma "0000
\mathchardef\Delta "0001
\mathchardef\Theta "0002
\mathchardef\Lambda "0003
\mathchardef\Xi "0004
\mathchardef\Pi "0005
\mathchardef\Sigma "0006
\mathchardef\Upsilon "0007
\mathchardef\Phi "0008
\mathchardef\Psi "0009
\mathchardef\Omega "000A
%
%
\newcount\firstpage  \firstpage=1  
\newcount\jnl                      
\newcount\secno                    
\newcount\subno                    
\newcount\subsubno                 
\newcount\appno                    
\newcount\tabno                    
\newcount\figno                    
\newcount\countno                  
\newcount\refno                    
\newcount\eqlett     \eqlett=97    
\newif\ifletter
\newif\ifwide
\newif\ifnotfull
\newif\ifaligned
\newif\ifnumbysec  
\newif\ifappendix
\newif\ifnumapp
\newif\ifssf
\newif\ifppt
\newdimen\t@bwidth
\newdimen\c@pwidth
\newdimen\digitwidth                    
\newdimen\argwidth                      
\newdimen\secindent    \secindent=5pc   
\newdimen\textind    \textind=16pt      
\newdimen\tempval                       
\newskip\beforesecskip
\def\beforesecspace{\vskip\beforesecskip\relax}
\newskip\beforesubskip
\def\beforesubspace{\vskip\beforesubskip\relax}
\newskip\beforesubsubskip
\def\beforesubsubspace{\vskip\beforesubsubskip\relax}
\newskip\secskip
\def\secspace{\vskip\secskip\relax}
\newskip\subskip
\def\subspace{\vskip\subskip\relax}
\newskip\insertskip
\def\insertspace{\vskip\insertskip\relax}
\def\sp@ce{\ifx\next*\let\next=\@ssf
               \else\let\next=\@nossf\fi\next}
\def\@ssf#1{\nobreak\secspace\global\ssftrue\nobreak}
\def\@nossf{\nobreak\secspace\nobreak\noindent\ignorespaces}
\def\subsp@ce{\ifx\next*\let\next=\@sssf
               \else\let\next=\@nosssf\fi\next}
\def\@sssf#1{\nobreak\subspace\global\ssftrue\nobreak}
\def\@nosssf{\nobreak\subspace\nobreak\noindent\ignorespaces}
\beforesecskip=24pt plus12pt minus8pt
\beforesubskip=12pt plus6pt minus4pt
\beforesubsubskip=12pt plus6pt minus4pt
\secskip=12pt plus 2pt minus 2pt
\subskip=6pt plus3pt minus2pt
\insertskip=18pt plus6pt minus6pt%
\fontdimen16\tensy=2.7pt
\fontdimen17\tensy=2.7pt
%
%
\def\eqlabel{(\ifappendix\applett
               \ifnumbysec\ifnum\secno>0 \the\secno\fi.\fi
               \else\ifnumbysec\the\secno.\fi\fi\the\countno)}
\def\seclabel{\ifappendix\ifnumapp\else\applett\fi
    \ifnum\secno>0 \the\secno
    \ifnumbysec\ifnum\subno>0.\the\subno\fi\fi\fi
    \else\the\secno\fi\ifnum\subno>0.\the\subno
         \ifnum\subsubno>0.\the\subsubno\fi\fi}
\def\tablabel{\ifappendix\applett\fi\the\tabno}
\def\figlabel{\ifappendix\applett\fi\the\figno}
\def\gac{\global\advance\countno by 1}
%
%

\def\vfootnote#1{\insert\footins\bgroup
\interlinepenalty=\interfootnotelinepenalty
\splittopskip=\ht\strutbox 
\splitmaxdepth=\dp\strutbox \floatingpenalty=20000
\leftskip=0pt \rightskip=0pt \spaceskip=0pt \xspaceskip=0pt%
\noindent\smallfonts\rm #1\ \ignorespaces\footstrut\futurelet\next\fo@t}
%
%
\def\endinsert{\egroup
    \if@mid \dimen@=\ht0 \advance\dimen@ by\dp0
       \advance\dimen@ by12\p@ \advance\dimen@ by\pagetotal
       \ifdim\dimen@>\pagegoal \@midfalse\p@gefalse\fi\fi
    \if@mid \insertspace \box0 \par \ifdim\lastskip<\insertskip
    \removelastskip \penalty-200 \insertspace \fi
    \else\insert\topins{\penalty100
       \splittopskip=0pt \splitmaxdepth=\maxdimen 
       \floatingpenalty=0
       \ifp@ge \dimen@=\dp0
       \vbox to\vsize{\unvbox0 \kern-\dimen@}%
       \else\box0\nobreak\insertspace\fi}\fi\endgroup}   
%
%
%
\def\ind{\hbox to \secindent{\hfill}}
%
%

%
%
\def\lo#1{\llap{${}#1{}$}}
%
%
\def\indeqn#1{\alignedfalse\displ@y\halign{\hbox to \displaywidth
    {$\ind\@lign\displaystyle##\hfil$}\crcr #1\crcr}}
%
%
\def\indalign#1{\alignedtrue\displ@y \tabskip=0pt 
  \halign to\displaywidth{\ind$\@lign\displaystyle{##}$\tabskip=0pt
    &$\@lign\displaystyle{{}##}$\hfill\tabskip=\centering
    &\llap{$\@lign\hbox{\rm##}$}\tabskip=0pt\crcr
    #1\crcr}}
\def\indenteddisplay#1$${\indispl@y{#1 }}
\def\indispl@y#1{\disptest#1\eqalignno\eqalignno\disptest}
\def\disptest#1\eqalignno#2\eqalignno#3\disptest{%
    \ifx#3\eqalignno
    \indalign#2%
    \else\indeqn{#1}\fi$$}
%
%

%
%

%
%

%
%

%
%

\def\ns{\noalign{\vskip-3pt}}

%

%
%
\def\bhbar{\rlap{\kern1pt\raise.4ex\hbox{\bf\char'40}}\bi{h}}

\def\frac#1#2{{#1\over#2}}
\ifams
\def\lap{\lesssim}
\def\gap{\gtrsim}

\let\leq=\leqslant

\let\geq=\geqslant
\else

\def\gap{\;\lower3pt\hbox{$\buildrel > \over \sim$}\;}%
\def\lap{\;\lower3pt\hbox{$\buildrel < \over \sim$}\;}\fi

\chardef\ii="10
\def\tqs{\hbox to 25pt{\hfil}}

\def\Bbbone{1\kern-.22em {\rm l}}
%
%
\def\rp{\raise8pt\hbox{$\scriptstyle\prime$}}
%
%
%
%

%
%
\def\[#1\]{\setbox0=\hbox{$\dsty#1$}\argwidth=\wd0
    \setbox0=\hbox{$\left[\box0\right]$}\advance\argwidth by -\wd0
    \left[\kern.3\argwidth\box0\kern.3\argwidth\right]}
%
%
\def\lsb#1\rsb{\setbox0=\hbox{$#1$}\argwidth=\wd0
    \setbox0=\hbox{$\left[\box0\right]$}\advance\argwidth by -\wd0
    \left[\kern.3\argwidth\box0\kern.3\argwidth\right]}
%

%
%

%
\def\pt(#1){({\it #1\/})}
\let\dsty=\displaystyle

%
%
\def\reactions#1{\vskip 12pt plus2pt minus2pt%
\vbox{\hbox{\kern\secindent\vrule\kern12pt%
\vbox{\kern0.5pt\vbox{\hsize=24pc\parindent=0pt\smallfonts\rm NUCLEAR 
REACTIONS\strut\quad #1\strut}\kern0.5pt}\kern12pt\vrule}}}
%
%
\def\slashchar#1{\setbox0=\hbox{$#1$}\dimen0=\wd0%
\setbox1=\hbox{/}\dimen1=\wd1%
\ifdim\dimen0>\dimen1%
\rlap{\hbox to \dimen0{\hfil/\hfil}}#1\else                                        
\rlap{\hbox to \dimen1{\hfil$#1$\hfil}}/\fi}
%
%
\def\textindent#1{\noindent\hbox to \parindent{#1\hss}\ignorespaces}
%
%
\def\opencirc{\raise1pt\hbox{$\scriptstyle{\bigcirc}$}}

\ifams
\def\opensqr{\hbox{$\square$}}

\def\opentridown{\hbox{$\triangledown$}}

\else
\def\opensqr{\vbox{\hrule height.4pt\hbox{\vrule width.4pt height3.5pt
    \kern3.5pt\vrule width.4pt}\hrule height.4pt}}

\def\opentridown{\raise1pt\hbox{$\scriptstyle\bigtriangledown$}}

\fi

%
%
\def\m@th{\mathsurround=0pt}
%
%
\def\cases#1{%
\left\{\,\vcenter{\normalbaselines\openup1\jot\m@th%
     \ialign{$\displaystyle##\hfil$&\rm\tqs##\hfil\crcr#1\crcr}}\right.}%
%
%
\def\oldcases#1{\left\{\,\vcenter{\normalbaselines\m@th
    \ialign{$##\hfil$&\rm\quad##\hfil\crcr#1\crcr}}\right.}
%
%
\def\numcases#1{\left\{\,\vcenter{\baselineskip=15pt\m@th%
     \ialign{$\displaystyle##\hfil$&\rm\tqs##\hfil
     \crcr#1\crcr}}\right.\hfill
     \vcenter{\baselineskip=15pt\m@th%
     \ialign{\rlap{$\phantom{\displaystyle##\hfil}$}\tabskip=0pt&\en
     \rlap{\phantom{##\hfil}}\crcr#1\crcr}}}
\def\ptnumcases#1{\left\{\,\vcenter{\baselineskip=15pt\m@th%
     \ialign{$\displaystyle##\hfil$&\rm\tqs##\hfil
     \crcr#1\crcr}}\right.\hfill
     \vcenter{\baselineskip=15pt\m@th%
     \ialign{\rlap{$\phantom{\displaystyle##\hfil}$}\tabskip=0pt&\enpt
     \rlap{\phantom{##\hfil}}\crcr#1\crcr}}\global\eqlett=97
     \global\advance\countno by 1}
%
%
\def\eq(#1){\ifaligned\@mp(#1)\else\hfill\llap{{\rm (#1)}}\fi}
\def\ceq(#1){\ns\ns\ifaligned\@mp\fi\eq(#1)\cr\ns\ns}
\def\eqpt(#1#2){\ifaligned\@mp(#1{\it #2\/})
                    \else\hfill\llap{{\rm (#1{\it #2\/})}}\fi}
\let\eqno=\eq
%
%
\countno=1

\def\aleq{&\rm(\ifappendix\applett
               \ifnumbysec\ifnum\secno>0 \the\secno\fi.\fi
               \else\ifnumbysec\the\secno.\fi\fi\the\countno}
\def\noaleq{\hfill\llap\bgroup\rm(\ifappendix\applett
               \ifnumbysec\ifnum\secno>0 \the\secno\fi.\fi
               \else\ifnumbysec\the\secno.\fi\fi\the\countno}
\def\@mp{&}
\def\en{\ifaligned\aleq)\else\noaleq)\egroup\fi\gac}
\def\cen{\ns\ns\ifaligned\@mp\fi\en\cr\ns\ns}
\def\enpt{\ifaligned\aleq{\it\char\the\eqlett})\else
    \noaleq{\it\char\the\eqlett})\egroup\fi
    \global\advance\eqlett by 1}
\def\endpt{\ifaligned\aleq{\it\char\the\eqlett})\else
    \noaleq{\it\char\the\eqlett})\egroup\fi
    \global\eqlett=97\gac}
%
%




%
%

\def\NP{{\it Nucl. Phys.}}
\def\PL{{\it Phys. Lett.}}
\def\PR{{\it Phys. Rev.}}

\def\ZP{{\it Z. Phys.}}
\headline={\ifodd\pageno{\ifnum\pageno=\firstpage\hfill
   \else\rrhead\fi}\else\lrhead\fi}
\def\rrhead{\textfonts\hskip\secindent\it
    \shorttitle\hfill\rm\folio}
\def\lrhead{\textfonts\hbox to\secindent{\rm\folio\hss}%
    \it\aunames\hss}
\footline={\ifnum\pageno=\firstpage \hfill\textfonts\rm\folio\fi}
\def\@rticle#1#2{\vglue.5pc
    {\parindent=\secindent \bf #1\par}
     \vskip2.5pc
    {\exhyphenpenalty=10000\hyphenpenalty=10000
     \baselineskip=18pt\raggedright\noindent
     \headfonts\bf#2\par}\futurelet\next\sh@rttitle}%
\def\title#1{\gdef\shorttitle{#1}
    \vglue4pc{\exhyphenpenalty=10000\hyphenpenalty=10000 
    \baselineskip=18pt 
    \raggedright\parindent=0pt
    \headfonts\bf#1\par}\futurelet\next\sh@rttitle} 

\def\article#1#2{\gdef\shorttitle{#2}\@rticle{#1}{#2}} 
\def\review#1{\gdef\shorttitle{#1}%
    \@rticle{REVIEW \ifpbm\else ARTICLE\fi}{#1}}
\def\topical#1{\gdef\shorttitle{#1}%
    \@rticle{TOPICAL REVIEW}{#1}}
\def\comment#1{\gdef\shorttitle{#1}%
    \@rticle{COMMENT}{#1}}
\def\note#1{\gdef\shorttitle{#1}%
    \@rticle{NOTE}{#1}}
\def\prelim#1{\gdef\shorttitle{#1}%
    \@rticle{PRELIMINARY COMMUNICATION}{#1}}
\def\letter#1{\gdef\shorttitle{Letter to the Editor}%
     \gdef\aunames{Letter to the Editor}
     \global\lettertrue\ifnum\jnl=7\global\letterfalse\fi
     \@rticle{LETTER TO THE EDITOR}{#1}}
\def\sh@rttitle{\ifx\next[\let\next=\sh@rt
                \else\let\next=\f@ll\fi\next}
\def\sh@rt[#1]{\gdef\shorttitle{#1}}
\def\f@ll{}
\def\author#1{\ifletter\else\gdef\aunames{#1}\fi\vskip1.5pc
    {\parindent=\secindent  
     \hang\textfonts  
     \ifppt\bf\else\rm\fi#1\par}  
     \ifppt\bigskip\else\smallskip\fi
     \futurelet\next\@unames}
\def\@unames{\ifx\next[\let\next=\short@uthor
                 \else\let\next=\@uthor\fi\next}
\def\short@uthor[#1]{\gdef\aunames{#1}}
\def\@uthor{}
\def\address#1{{\parindent=\secindent
    \exhyphenpenalty=10000\hyphenpenalty=10000
\ifppt\textfonts\else\smallfonts\fi\hang\raggedright\rm#1\par}%
    \ifppt\bigskip\fi}
\def\jl#1{\global\jnl=#1}
\jl{0}%
\def\journal{\ifnum\jnl=1 J. Phys.\ A: Math.\ Gen.\ 
        \else\ifnum\jnl=2 J. Phys.\ B: At.\ Mol.\ Opt.\ Phys.\ 
        \else\ifnum\jnl=3 J. Phys.:\ Condens. Matter\ 
        \else\ifnum\jnl=4 J. Phys.\ G: Nucl.\ Part.\ Phys.\ 
        \else\ifnum\jnl=5 Inverse Problems\ 
        \else\ifnum\jnl=6 Class. Quantum Grav.\ 
        \else\ifnum\jnl=7 Network\ 
        \else\ifnum\jnl=8 Nonlinearity\
        \else\ifnum\jnl=9 Quantum Opt.\
        \else\ifnum\jnl=10 Waves in Random Media\
        \else\ifnum\jnl=11 Pure Appl. Opt.\ 
        \else\ifnum\jnl=12 Phys. Med. Biol.\
        \else\ifnum\jnl=13 Modelling Simulation Mater.\ Sci.\ Eng.\ 
        \else\ifnum\jnl=14 Plasma Phys. Control. Fusion\ 
        \else\ifnum\jnl=15 Physiol. Meas.\ 
        \else\ifnum\jnl=16 Sov.\ Lightwave Commun.\
        \else\ifnum\jnl=17 J. Phys.\ D: Appl.\ Phys.\
        \else\ifnum\jnl=18 Supercond.\ Sci.\ Technol.\
        \else\ifnum\jnl=19 Semicond.\ Sci.\ Technol.\
        \else\ifnum\jnl=20 Nanotechnology\
        \else\ifnum\jnl=21 Meas.\ Sci.\ Technol.\ 
        \else\ifnum\jnl=22 Plasma Sources Sci.\ Technol.\ 
        \else\ifnum\jnl=23 Smart Mater.\ Struct.\ 
        \else\ifnum\jnl=24 J.\ Micromech.\ Microeng.\
   \else Institute of Physics Publishing\ 
   \fi\fi\fi\fi\fi\fi\fi\fi\fi\fi\fi\fi\fi\fi\fi
   \fi\fi\fi\fi\fi\fi\fi\fi\fi}
\let\abs=\beginabstract

\let\endabs=\endabstract
\def\submitted{\ifppt\noindent\textfonts\rm Submitted to \journal\par
     \bigskip\fi}
\def\today{\number\day\ \ifcase\month\or
     January\or February\or March\or April\or May\or June\or
     July\or August\or September\or October\or November\or
     December\fi\space \number\year}
\def\date{\ifppt\noindent\textfonts\rm 
     Date: \today\par\goodbreak\bigskip\fi}
%
%
\def\pacs#1{\ifppt\noindent\textfonts\rm 
     PACS number(s): #1\par\bigskip\fi}
%

%
%
\def\section#1{\ifppt\ifnum\secno=0\eject\fi\fi
    \subno=0\subsubno=0\global\advance\secno by 1
    \gdef\labeltype{\seclabel}\ifnumbysec\countno=1\fi
    \goodbreak\beforesecspace\nobreak
    \noindent{\bf \the\secno. #1}\par\futurelet\next\sp@ce}
\def\subsection#1{\subsubno=0\global\advance\subno by 1
     \gdef\labeltype{\seclabel}%
     \ifssf\else\goodbreak\beforesubspace\fi
     \global\ssffalse\nobreak
     \noindent{\it \the\secno.\the\subno. #1\par}%
     \futurelet\next\subsp@ce}
\def\subsubsection#1{\global\advance\subsubno by 1
     \gdef\labeltype{\seclabel}%
     \ifssf\else\goodbreak\beforesubsubspace\fi
     \global\ssffalse\nobreak
     \noindent{\it \the\secno.\the\subno.\the\subsubno. #1}\null. 
     \ignorespaces}
%

%
%
\def\numappendix#1{\ifappendix\ifnumbysec\countno=1\fi\else
    \countno=1\figno=0\tabno=0\fi
    \subno=0\global\advance\appno by 1
    \secno=\appno\gdef\applett{A}\gdef\labeltype{\seclabel}%
    \global\appendixtrue\global\numapptrue
    \goodbreak\beforesecspace\nobreak
    \noindent{\bf Appendix \the\appno. #1\par}%
    \futurelet\next\sp@ce}
\def\numsubappendix#1{\global\advance\subno by 1\subsubno=0
    \gdef\labeltype{\seclabel}%
    \ifssf\else\goodbreak\beforesubspace\fi
    \global\ssffalse\nobreak
    \noindent{\it A\the\appno.\the\subno. #1\par}%
    \futurelet\next\subsp@ce}
\def\@ppendix#1#2#3{\countno=1\subno=0\subsubno=0\secno=0\figno=0\tabno=0
    \gdef\applett{#1}\gdef\labeltype{\seclabel}\global\appendixtrue
    \goodbreak\beforesecspace\nobreak
    \noindent{\bf Appendix#2#3\par}\futurelet\next\sp@ce}
\def\Appendix#1{\@ppendix{A}{. }{#1}}
\def\appendix#1#2{\@ppendix{#1}{ #1. }{#2}}
\def\App#1{\@ppendix{A}{ }{#1}}
\def\app{\@ppendix{A}{}{}}
\def\subappendix#1#2{\global\advance\subno by 1\subsubno=0
    \gdef\labeltype{\seclabel}%
    \ifssf\else\goodbreak\beforesubspace\fi
    \global\ssffalse\nobreak
    \noindent{\it #1\the\subno. #2\par}%
    \nobreak\subspace\noindent\ignorespaces}
%
%
\def\@ck#1{\ifletter\bigskip\noindent\ignorespaces\else
    \goodbreak\beforesecspace\nobreak
    \noindent{\bf Acknowledgment#1\par}%
    \nobreak\secspace\noindent\ignorespaces\fi}
\def\ack{\@ck{s}}
\def\ackn{\@ck{}}
\def\n@ip#1{\goodbreak\beforesecspace\nobreak
    \noindent\smallfonts{\it #1}. \rm\ignorespaces}
\def\naip{\n@ip{Note added in proof}}
\def\na{\n@ip{Note added}}

%
%

%

%
%

%

%

\def\tablecont{\topinsert\global\advance\tabno by -1
    \tablecaption{(continued)}}
\def\tablecaption#1{\gdef\labeltype{\tablabel}\global\widefalse
    \leftskip=\secindent\parindent=0pt
    \global\advance\tabno by 1
    \smallfonts{\bf Table \ifappendix\applett\fi\the\tabno.} \rm #1\par
    \smallskip\futurelet\next\t@b}
\def\t@b{\ifx\next*\let\next=\widet@b
             \else\ifx\next[\let\next=\fullwidet@b
                      \else\let\next=\narrowt@b\fi\fi
             \next}
\def\widet@b#1{\global\widetrue\global\notfulltrue
    \t@bwidth=\hsize\advance\t@bwidth by -\secindent} 
\def\fullwidet@b[#1]{\global\widetrue\global\notfullfalse
    \leftskip=0pt\t@bwidth=\hsize}                  
\def\narrowt@b{\global\notfulltrue}
\def\align{\catcode`?=13\ifnotfull\moveright\secindent\fi
    \vbox\bgroup\halign\ifwide to \t@bwidth\fi
    \bgroup\strut\tabskip=1.2pc plus1pc minus.5pc}
\def\endalign{\egroup\egroup\catcode`?=12}

%
%

%
%

%

%
%

%

\catcode`?=13
\def\lineup{\setbox0=\hbox{\smallfonts\rm 0}%
    \digitwidth=\wd0%
    \def?{\kern\digitwidth}%
    \def\\{\hbox{$\phantom{-}$}}%
    \def\-{\llap{$-$}}}
\catcode`?=12
%
%
\def\sidetable#1#2{\hbox{\ifppt\hsize=18pc\t@bwidth=18pc
                          \else\hsize=15pc\t@bwidth=15pc\fi
    \parindent=0pt\vtop{\null #1\par}%
    \ifppt\hskip1.2pc\else\hskip1pc\fi
    \vtop{\null #2\par}}} 
\def\lstable#1#2{\everypar{}\tempval=\hsize\hsize=\vsize
    \vsize=\tempval\hoffset=-3pc
    \global\tabno=#1\gdef\labeltype{\tablabel}%
    \noindent\smallfonts{\bf Table \ifappendix\applett\fi
    \the\tabno.} \rm #2\par
    \smallskip\futurelet\next\t@b}
\def\inctabno{\global\advance\tabno by 1}
%
%
\def\Figures{\vfill\eject\global\appendixfalse\textfonts\rm
    \everypar{}\noindent{\bf Figure captions}\par
    \bigskip}
\def\figure#1{\figc@ption{#1}\bigskip}
\def\figc@ption#1{\global\advance\figno by 1\gdef\labeltype{\figlabel}%
   {\parindent=\secindent\smallfonts\hang
    {\bf Figure \ifappendix\applett\fi\the\figno.} \rm #1\par}}
%
%
\def\refHEAD{\goodbreak\beforesecspace
     \noindent\textfonts{\bf References}\par
     \let\ref=\rf
     \nobreak\smallfonts\rm}
\def\numreferences{\refHEAD\parindent=30pt
     \everypar{\hang\noindent\frenchspacing\rm}
     \secspace}
\def\rf#1{\par\noindent\hbox to 21pt{\hss #1\quad}\ignorespaces}
%

%

%
%
\def\numrefjl#1#2#3#4#5{\par\rf{#1}#2 {\it #3 \bf #4} #5\par}
%
%

%
%

%
%

%
\catcode`\@=12
%
%

%
%
\def\jnlstyle{\pptfalse\headsize{14}{18}%
\textsize{10}{12}%
\smallsize{8}{10}
\textind=16pt}
%
%

%
%

%
\parindent=\textind


%
\chardef\other=12
\def\deactivate{%
  \catcode`\\=\other \catcode`\{=\other
  \catcode`\}=\other \catcode`\$=\other
  \catcode`\&=\other \catcode`\#=\other
  \catcode`\%=\other \catcode`\~=\other
  \catcode`\^=\other \catcode`\_=\other}

{\obeylines\gdef\startdisplay#1
  {\catcode`\^^M=5$$#1\halign\bgroup\indent##\hfil&&\qquad##\hfil\cr}}
\outer\def\enddisplay{\crcr\egroup$$}
\def\ttverbatim{\begingroup\deactivate\obeyspaces\obeylines \tt}
{\obeyspaces\gdef {\ }}  
\catcode`\|=\active
{\obeylines
\gdef|{\ttverbatim\spaceskip=.5em plus.25em minus.15em%
\let^^M=\ \let|=\endgroup}}%

%
%
\def\makeactive#1{\catcode`#1=\active\ignorespaces}
{
  \makeactive\^^M %
  \gdef\obeywhitespace{%
    \makeactive\^^M %
    \let^^M=\newline %
    \aftergroup\removebox 
    \obeyspaces %
  }%
}
\def\newline{\par\noindent}
\def\removebox{\setbox0=\lastbox}
\def\|{|}


%
%
\catcode`@=11
\def\LaTeX{L\kern-.26em \raise.6ex\hbox{\csname\vsTEXT rm\@nd A}%
   \kern-.15em\TeX}%
\def\AmSTeX{{$\cal{A}$}\kern-.1667em\lower.5ex\hbox{$\cal{M}$}%
   \kern-.125em{$\cal{S}$}-\TeX}
\catcode`@=12

%
%

%
\jnlstyle          

\title{Heavy Quarks in DIS (Theory)}

\author{R S Thorne\dag}[R S Thorne]

\address{\dag\ Theoretical Physics, Department of Physics, 1 Keble Road,
Oxford OX1 3NP, UK}

\abs
I review the various methods for taking account of finite quark masses in
DIS and related processes. I pay particular attention to the so-called
variable flavour number schemes (VFNS) which are designed to 
extrapolate smoothly
from the region near threshold of production of heavy quark pairs to the
region where the quarks become effectively massless.
\endabs

\pacs{0000}

\submitted

\date
   
\vskip 0.5in

\centerline{Plenary talk given at}
\centerline{3rd UK Phenomenology Workshop on HERA Physics,}
\centerline{Durham, UK, 20-25 September 1998.}
\centerline{{\it To appear in the proceedings.}} 

\vskip 0.3in

\section{Introduction}

In the past  few years there has been a considerable effort to improve the
theoretical description of the effects due to heavy quarks (i.e $m_H\gg
\Lambda_{QCD}$) in DIS and related scattering processes. As is often the
case this has been driven by the vast improvement in experimental data
rendering the previous simplistic theoretical description inadequate. There is
now not only data on the charm component of the structure function at
relatively high $x$ ($x>0.01$) from the EMC collaboration [1], but also at 
very low $x$ ($0.00005<x<0.01$) from HERA [2,3]. This latter data also exists 
not only in
terms of the total inclusive cross-section, but also in more differential
forms, which is presented by J Cole [4] in the experimental plenary talk. 
Moreover,
the charm component to the total structure function now comprises about $25\%$
of the total at the highest $y$ values, and with the extremely precise data
now available a more sophisticated treatment of heavy quark effects is
needed here. Perhaps surprisingly, this is also the case when comparing the
theoretical predictions to the NMC data [5] at $x \sim 0.05$ and 
$Q^2\sim 10 {\hbox{\rm GeV}}^2$,
since although charm only constitutes $5-10\%$ of the total structure function
it contributes more strongly to the value of $d F_2/d\ln Q^2$, which is the
quantity we are really testing in fits to data.

\section{Theoretical Methods}

The oldest, and by far the simplest method for treating heavy flavors in
DIS (and other processes) is known as the zero mass variable flavour scheme
(ZM-VFNS).
Using this method one treats the heavy quark as being infinitely massive
below some scale $\mu_T$, i.e. it decouples from the theory below this scale, 
and as completely
massless above $\mu_T$. Therefore the number of quarks flavors just
changes from $n_f$ to $n_{f+1}$ in the massless expressions for splitting
functions, coefficient functions etc. at some scale usually chosen as
$\mu_T^2=m_H^2$. It seems intuitively obvious that if the evolution of the
massive quark starts at $\mu^2=m_H^2$, it should evolve from a value of
roughly zero at this scale. However, this is hardly quantitative, and the 
real content of the ZM-VFNS is the set of matching conditions required to 
guarantee that the correct results are obtained for both
$Q^2/m_H^2 \to 0$ and $Q^2/m_H^2 \to \infty$ [6]. This results in a 
prescription
for how the coupling should change at the matching point, and also for the
form of the parton distributions at this point. In fact one can calculate 
the partons in the massless $n_f+1$ flavour scheme entirely 
in terms of those in the $n_f$ flavour scheme, i.e. 
$$
f^{n_f+1}_k(\mu^2,m_H^2)=A_{kj}(\mu^2/m_H^2)\otimes f^{n_f}_j(\mu^2),
\eqno(1)
$$
where the matrix elements are perturbatively calculable, and can be found to 
${\cal O}(\alpha_s^2)$ in [7]. 
This expression is correct up to higher twist (${\cal O}(\Lambda_{QCD}^2/
m_H^2)$) corrections due to the intrinsic heavy quark distribution. 
It turns out that if 
we do choose precisely $\mu^2_T=m_H^2$ then to NLO the heavy quark distribution
does start form zero. AT NNLO, i.e. ${\cal O}(\alpha_s^2)$, this is no longer
true, but the non-zero starting distribution is known [7]. 
The ZM-VFNS has been used in global fits to structure function data for many
years. However, it has a clear deficiency, i.e. the threshold treatment. 
It takes no account of the fact
that quark pairs can only be created for $W^2=Q^2(1/x-1)\geq 4m_H^2$, and
indeed has an error of ${\cal O}(m_H^2/Q^2)$ reflecting this. So until
we reach $Q^2\gg m_H^2$, it is not quantitatively useful, and is clearly not
sufficient to describe much of the data now available.

In some senses the opposite to the ZM-VFNS is the fixed flavour number scheme
(FFNS). In this scheme only those quarks with $m_H\leq \Lambda_{QCD}$ are 
treated as partons. All heavy quark production is generated via the
perturbatively calculable coefficient functions
describing the scattering of initial light partons. The finite quark mass
effects are kept in the calculation of these coefficient functions, and hence
the threshold behaviour for quark pair production is described correctly.
The error on the structure function calculations in this scheme is
${\cal O}(\Lambda_{QCD}^2/m_H^2)$, reflecting the decoupling of the 
intrinsic heavy quark
effects from the light parton distributions. This prescription thus works
very well near the threshold region.  Its only real potential problem
is in the region of high $Q^2$. This is because the perturbative series for the
coefficient functions behaves like (letting $\mu^2=Q^2$ from now on)
$$
C(\alpha_s(Q^2), m_H^2/Q^2) = \sum_{n=1}^{\infty}
\sum_{p=q}^{n}\alpha_s^n(Q^2)\ln^{n-p}
(Q^2/m_H^2), \eqno(2)
$$
when $Q^2/m_H^2 \to \infty$, and $q$ is a process dependent number ($q=0$ for
$F_2(x,Q^2)$ but $q=1$ for $F_L(x,Q^2)$). Thus, a fixed order in $\alpha_s$
calculation is not guaranteed to give a good representation of the correct
result at high $Q^2$, and it would be desirable to sum the series in leading
powers of $\ln(Q^2/m_H^2)$. This is precisely what is done in actually
treating the heavy quark as a parton in the ZM-VFNS, and determining its
evolution by solving the renormalization group equations. Hence, although the
FFNS seems preferable to the ZM-VFNS, it would seem, at least in principle,
desirable to devise a scheme which contains the best parts of both the
ZM-VFNS and the FFNS. Indeed, this is necessary if one actually wishes to
consider parton distributions for heavy flavors without resigning
oneself to large errors near threshold.

Various attempts have been made to achieve this objective, and the procedures
are generally known as variable flavour number schemes. This is because
they all have in common the features that for low scales, usually $Q^2\leq
m_H^2$, they are identical to the FFNS, having 3 partonic quarks, while for
$Q^2/m_H^2 \to \infty$ the heavy quark is treated as a massless parton,
evolving according to the massless evolution equations. There are at least
4 different versions of a VFNS currently available. One, of these, devised by
Buza {\it et al}. [8], is at the level of structure functions rather than
parton distributions, extrapolating smoothly from $F^{FFNS}(x,Q^2)$ at low 
$Q^2$ to $F^{ZM-VFNS}(x,Q^2)$ at high $Q^2$ via the subtraction of 
$F^{ASYM}(x,Q^2)$ (the FFNS result
with all ${\cal O}(m_H^2/Q^2)$ corrections removed), which 
is the limit of the
other two forms as $Q^2/m_H^2 \to 1$ or $\to \infty$. Another, the 
MRRS scheme [9],
involves  mass corrections to the evolution of the partons. Despite the fact
that these disappear as $Q^2/m_H^2 \to \infty$, their presence in the evolution
means that neither the parton distributions or the coefficient functions are
precisely as they would be in the ZM-VFNS, even at asymptotic $Q^2$. In this
review I will concentrate on those schemes where the treatment of the
parton distributions is exactly as in the ZM-VFNS, and hence where the
theories tend unambiguously to this asymptotic limit.

The first of these schemes is known as the ACOT scheme [10]. It is defined by
calculating the coefficient functions with both
light and heavy initial quarks. The divergences in logs of $(Q^2/m_H^2)$ 
are then
factored in the same manner as the infrared divergences due to  
light initial states, and
the heavy parton distributions defined via this procedure are guaranteed to
obey the massless evolution equations, i.e. they are precisely 
the parton distributions
defined in eq. (1). The coefficient functions remaining are 
guaranteed to be finite, and in fact 
identical to the massless coefficient functions in the limit $Q^2/m_H^2 \to
\infty$. However, they contain all the mass effects important at lower $Q^2$. 
At all orders this method will give exactly the same results as the FFNS, but
the way of ordering the expansion is now somewhat different.
The coefficient functions are
related to those in the FFNS by this exact equality of the 
structure functions calculated via the two methods:
$$
F(x,Q^2)=C^{FFNS}_k(Q^2/m_H^2)\otimes f^{n_f}_k(Q^2)\cr
\lo = C^{VFNS}_j(Q^2/m_H^2)\otimes A_{jk}(Q^2/m_H^2)
\otimes f^{n_f}_k(Q^2), \eqno(3)
$$
i.e. the $A_{jk}(Q^2/m_H^2)$ are the parts subtracted out of the divergent
coefficient functions which depend on the massive quarks. That this procedure 
is well-defined to all orders was recently formally proved by Collins [11]. 
This scheme has a well-defined method of calculating the mass-dependent 
coefficient functions in terms of diagrams with initial massive quarks
(for computational simplicity it is suggested in [11] that coefficient
functions with initial heavy quarks be calculated in the massless limit)
and the unique matrix elements $A_{jk}(Q^2/m_H^2)$. 
The only potential weakness of the approach is 
that coefficient functions now exist describing
the scattering of a single heavy 
quark. Such coefficient functions, e.g. the zeroth order $c+\gamma^{*} \to
c$, allow the creation of final state heavy quarks for $W^2<4m_H^2$, 
and hence the correct kinematical 
threshold is not respected order by order in the coefficient functions. The
factorization procedure guarantees that the correct threshold behaviour is
regained, but this is due to cancellations between orders,
and near threshold these cancellations can be larger than the final result,
as seen in [10].
Hence the ordering of terms in the ACOT scheme must be different to 
that for the calculation of light quark structure functions in order to 
obtain the most accurate results. 

More recently a different scheme has been proposed [12]. This takes the formal
result of eq.(3), and notes that since there is one more VFNS coefficient
function than there are FFNS coefficient functions, the former are not defined 
unambiguously. Therefore, it is decided to define the VFNS coefficient 
functions not from
the diagrammatic representation, but by solving eq.(3) subject to the 
physical constraint that the derivative of the structure function is 
continuous (in the very dominant gluon sector) across the matching point 
at $Q^2=m_H^2$. This results in heavy flavour coefficient functions which
are related to the $\ln Q^2$ derivatives of
$C^{FFNS,n}_g(x,Q^2/m_H^2)$ via 
the 
formal inverse of the LO quark-gluon splitting function, e.g. at leading order
$$
C^{VFNS,0}_{2,HH}(Q^2/m_H^2)\otimes p^0_{qg}={\partial C^{FFNS,1}_{2,g}
(x,Q^2/m_H^2) \over \partial \ln Q^2}. \eqno(4)
$$
Hence, at the loss of a
clear diagrammatic interpretation for  the coefficient functions one gains 
the correct type of threshold dependence in all coefficient functions. 
In this scheme the ordering of the calculation is the same as for light 
partons. To all orders the two VFNSs are identical, and have identical 
prescriptions for the partons at each order. They are effectively related by
a scheme change for coefficient functions which does not alter the parton 
distributions. There are in principle an infinite number of choices for
such schemes. 
  
However, the ambiguity in the VFNS coefficient functions only strictly
occurs because we are working in the limit where the 
$n_f+1$ flavour partons are completely determined by the $n_f$ flavour
distributions. (This is assumed in most practical uses of either of the VFNSs
in global fits.) This assumption leads to a minimum error of ${\cal O}
(\Lambda_{QCD}^2/m_H^2)$ due to the neglect of the effects of the 
intrinsic quark distribution (which though formally small, may be enhanced at 
large $x$ [13]) and its influence on the other parton
distributions. Collins has shown that within the ACOT scheme if one takes 
account of such effects, i.e. allows deviation from eq.(1) of this order,
then the calculation of the structure functions becomes accurate to 
${\cal O}(\Lambda_{QCD}^2/Q^2)$ [11]. Allowing this intrinsic
heavy quark distribution formally removes the 
redundancy in the definition of the coefficient functions, and the FFNS
and the alternative VFNS will now longer be identical to the ACOT scheme
if summed to all orders. 
In the FFNS a similar accuracy can only be obtained by adding 
intrinsic heavy quark effects in some {\it ad hoc}
manner. However, the accuracy in 
the alternative VFNS is in fact automatically of the same order
as the ACOT scheme. To appreciate this, let the parton
distributions , including the intrinsic heavy quarks, be identical
in the two schemes. The coefficient functions at 
each order are different by ${\cal O}(m_H^2/Q^2)$. Nevertheless,
this difference has been constructed to be unimportant
when combining with the part of the parton distributions 
insensitive to the intrinsic heavy quarks. The remainder of the parton 
distributions is of ${\cal O}
(\Lambda_{QCD}^2/m_H^2)$, and when combined with the difference in 
coefficient functions leads to a difference in structure functions of 
${\cal O}(\Lambda_{QCD}^2/Q^2)$, i.e. the same order as the 
accuracy of the ACOT scheme. Again, this will be true for the whole family 
of schemes
discussed above. Hence all VFNSs have in principle the same type of accuracy,
and none is guaranteed to be theoretically preferable.

\section{Results}

A comparison of the neutral current charm structure function obtained using 
the ZM-VFNS, the 
FFNS and the VFNS using the same input partons, and all at NLO is shown in
fig.1. The biggest difference is between the ZM-VFNS and the others at 
small $Q^2$, as we might expect. The differences between the VFNS and 
FFNS can be up to $\sim 20\%$. The failure of the ZM-VFNS at low $Q^2$ should 
surely mean that
it is no longer used, except when its defects are not important. The
FFNS is still used in many circumstances, including the GRV 
parameterizations [14],
and it is clear that for the neutral current structure functions 
its use does not lead to any obvious 
problems. 
Indeed, GRV show that the variation due to factorization/renormalization 
scale changes is similar to that seen between the different 
schemes in fig.1. However, 
despite this evidence the FFNS does have  potential shortcomings at high $Q^2$.
Indeed, it has been demonstrated in [15] that the charged current structure 
functions $F^{e^{+(-)}p}_{3,c}(x,Q^2)$  require the 
summation of large logs in ($Q^2/m_H^2)$ in order to obtain a stable result.
This may well be true for other processes, and the universality of parton 
distributions may be compromised in this scheme. 
  
For the moment most of the relevant experimental data is for the
neutral current structure function $F_2(x,Q^2)$. The direct data on the 
charm structure function is not 
yet accurate
enough to discriminate between the approaches,
and all work quite well (except for the ZM-VFNS for $Q^2\sim m_c^2$). 
However, the large amount of 
accurate data on the total structure function  does seem to be a discriminant.
In both [12] and a CTEQ fit [16] it was found that the respective VFNS fits 
were
superior to either the ZM-VFNS or the FFNS, with the latter faring worst.   
The improvement is seen not only for the HERA data but also for the NMC data.
It is difficult to make a qualitative comparison with the GRV FFNS 
calculation since they do not really perform a best fit, and hence produce
no $\chi^2$ values. Each of the above procedures is also easily applicable 
to the case of charged current scattering as well as neutral current, and 
as a further test it
would be interesting to see whether they result in any significant
difference for the strange component of the sea quarks extracted by the
CCFR collaboration [17].  

Hence, it is clear that there are a variety of different methods for dealing 
with heavy flavors in DIS. Each have their own virtues and there 
presently seems to be no overwhelming reason for choosing one exclusively,
though I would argue that a VFNS is the most sound theoretically. In
the working group session it was also demonstrated that each scheme could be 
subject to relatively important further corrections in the form of 
summation of large threshold logarithms [18]. Further quantitative analysis
of such effects using the techniques developed would lead to an even 
better understanding of heavy flavour production in DIS and supplement the 
significant progress made in this field in recent years. Also, the 
theoretical treatment is still a little behind the experimental development, 
and it would be useful to generalize the VFNS approaches to the 
situations where
more differential inclusive cross-sections, such as the $p_T$- or rapidity 
distributions of the final state heavy quarks are described, and where at 
present there are only massless and FFNS [19] results. Hence, I 
conclude that in the theoretical treatment of heavy
flavour physics for DIS significant progress has recently been made,
but there are many topics still to be investigated before we can claim to
have a really detailed understanding.    

\ack I would like to thank R.G. Roberts for continued collaboration on this 
topic, and J.C. Collins for helpful and detailed communications. I would
also like to thank M. Kr\"amer, E. Laenen,
S. Moch and  W. van Neerven for useful discussions.  

\numreferences

\numrefjl{[1]}{EMC collaboration: J. J. Aubert {\it et al.} 1983}{\NP}{B213}
{31}

\numrefjl{[2]}{H1 collaboration: C. Adloff {\it et al}. 1996}{\ZP}{C72}{593}

\numrefjl{[3]}{Zeus collaboration: J. Breitweg {\it et al}. 1997}{\PL}{B407}
{402}

\rf{[4]}J. Cole, these proceedings

\numrefjl{[5]}{NMC collaboration: M. Arneodo {\it et al}. 1997}{\NP}{B483}
{3}

\numrefjl{[6]}{J.C. Collins and W.K. Tung 1986}{\NP}{B278}{934}

\numrefjl{[7]}{M. Buza {\it et al}. 1996}{\NP}{B472}{611}
\numrefjl{}{M. Buza {\it et al}. 1998}{{\it Eur. Phys. J.}}{C1}{301}

\numrefjl{[8]}{M. Buza {\it et al}. 1997}{\PL}{B411}{211}

\numrefjl{[9]}{A.D. Martin {\it et al}. 1998}{{\it Eur. Phys. J.}}{C2}{287}

\numrefjl{[10]}{F. Olness and W.K. Tung 1988}{\NP}{B308}{813}
\numrefjl{}{M. Aivaziz, F. Olness and W. K. Tung 1994}{\PR}{D50}{3085}
\numrefjl{}{M. Aivaziz, J.C. Collins, F. Olness and W.K. Tung 1994}{\PR}
{D50}{3102}

\numrefjl{[11]}{J.C. Collins 1998}{\PR}{D58}{094002}

\numrefjl{[12]}{R.S. Thorne and R.G. Roberts 1998}{\PR}{D57}{6871}
\numrefjl{}{R.S. Thorne and R.G. Roberts 1998}{\PL}{B421}{303}

\rf{[13]} S.J. Brodsky, P. Hoyer, A.H. Mueller and W.K. Tang 1992
{\it Nucl. Phys.}{\bf B369} 519, and refs therein

\numrefjl{[14]}{M. Gl\"uck, E. Reya and A. Vogt 1998}{{\it Eur. Phys. J.}}
{C5}{461}

\numrefjl{[15]}{M. Buza and W.L. van Neerven 1997}{\NP}{B500}{301}

\numrefjl{[16]}{H.L. Lai and W.K. Tung 1997}{\ZP}{C74}{463}

\numrefjl{[17]}{CCFR collaboration: A.O. Bazarko {\it et al}. 1995}{\ZP}{C65}
{189}

\numrefjl{[18]}{E. Laenen and S.O. Moch 1999}{\PR}{D59}{034027}

\numrefjl{[19]}{B.W. Harris and J. Smith 1995}{\NP}{B452}{109}
\numrefjl{}{B.W. Harris and J. Smith 1995}{\PL}{B353}{535}

\Figures
 
\figure{Charm quark structure function, $F_{2,c}(x,Q^2)$ for $x=0.05$ and
$x=0.005$ calculated using the NLO VFNS (solid line) FFNS (dotted line) and 
ZM-VFNS (dashed line) with the same parton distributions and scale choice
($\mu^2=Q^2$) in each case.}

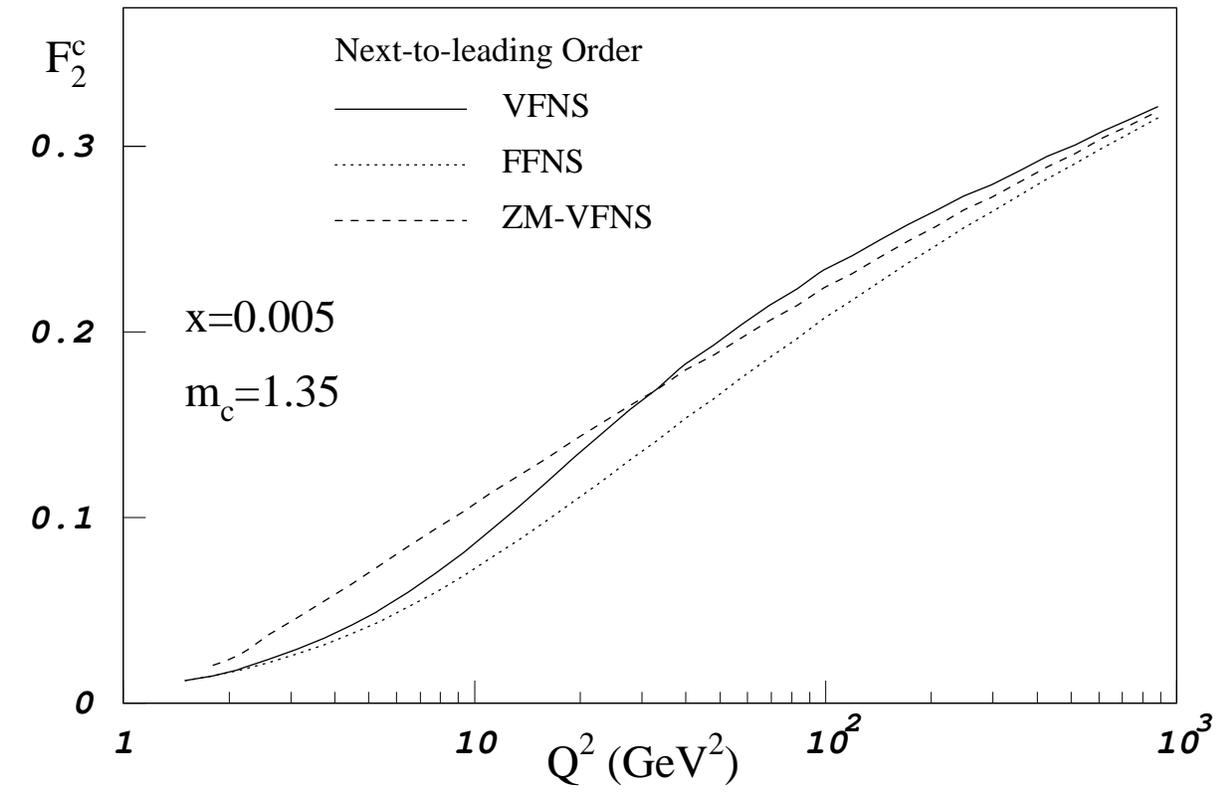
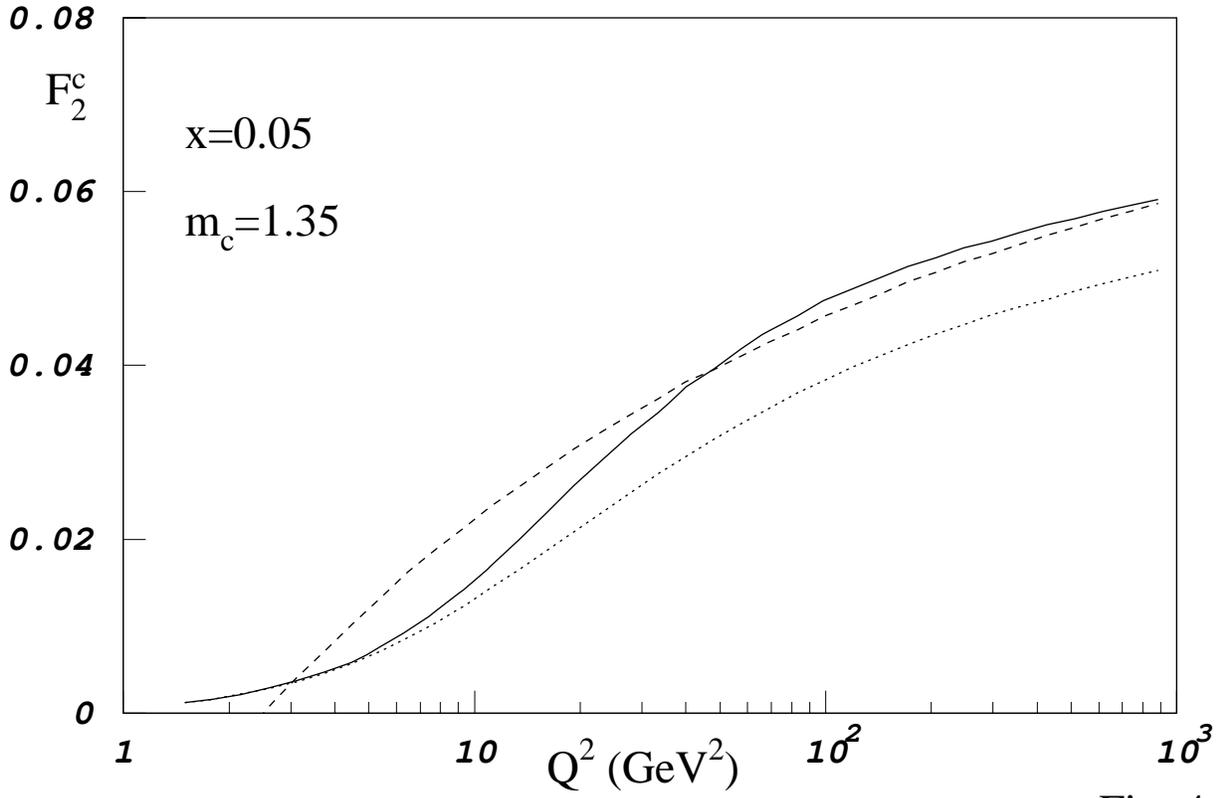
\bye